\documentclass[aps,prb,twocolumn,showpacs,superscriptaddress]{revtex4}

\usepackage{graphicx} 

\begin{document}

\title
{First principles calculations of spin-dependent conductance of graphene flakes}

\author{H. {\c S}ahin}
\affiliation{UNAM-Institute of Materials Science and
Nanotechnology, Bilkent University, 06800 Ankara, Turkey}

\author{R. T. Senger}
\email{senger@fen.bilkent.edu.tr} \affiliation{Department of
Physics, Bilkent University, 06800 Ankara, Turkey}

\date{\today}

\begin{abstract}

Using \textit{ab initio} density functional theory and quantum
transport calculations based on nonequilibrium Green's function
formalism we study structural, electronic, and transport
properties of hydrogen-terminated short graphene nanoribbons
(graphene flakes) and their functionalization with vanadium atoms.
Rectangular graphene flakes are stable, having geometric and
electronic structures quite similar to that of extended graphene
nanoribbons. We show that a spin-polarized current can be produced
by pure, hydrogenated rectangular graphene flakes by exploiting
the spatially-separated edge states of the flake using asymmetric,
non-magnetic contacts. Functionalization of the graphene flake
with magnetic adatoms such as vanadium also leads to
spin-polarized currents even with symmetric contacts. We observe
and discuss sharp discontinuities in the transmission spectra
which arise from Fano resonances of localized states in the flake.
\end{abstract}

\pacs{73.63.-b, 72.25.-b, 75.75.+a}


\maketitle

\section{Introduction}

In the last two decades various forms of carbon nanostructures,
namely buckyballs, carbon nanotubes, and lately graphene have
attracted a great deal of interest due to their novel fundamental
properties and possible applications in electronics. With the
developments in preparation and synthesis techniques carbon-based
nanostructures have emerged as one of the most promising materials
for non-silicon electronics.

The fabrication of graphene sheets\cite{novo1} and observation of
their unusual properties such as a half-integer quantum-Hall
effect have attracted much interest in electronic transport
properties of this type of two dimensional graphitic materials.
Observed gapless energy spectrum and high mobility electron
transport\cite{novo2,zhang, berger} are the most remarkable
features of graphene. It was shown by tight-binding calculations
considering the $\pi$ bands that in the electronic energy
dispersion of graphene, energy is linearly dependent on the wave
vector around the Fermi level\cite{wallace} which makes it a
unique material.

In recent experimental studies graphene nanoribbons (GNRs) with
narrow widths (10-70 nm) have been realized.\cite{han,li} Li
\textit{et al} report producing ultra narrow ribbons with widths
down to a few nanometers.\cite{li} In addition to high carrier
mobilities that are higher than those in commercial silicon
wafers, existence of width-dependent energy band gaps makes the
graphene nanoribbons a potentially useful structure for various
applications. The width dependence of the band gap and transport
properties in quasi-one dimensional narrow GNRs have been studied
theoretically.\cite{son,barone,areshkin,sasaki,abanin,lee,miyamoto,cresti,kudin,ezawa}

Graphene nanoribbons having $r$ zigzag rows ($r$-ZGNRs) are
predicted as semiconductors having a narrowing band gap with the
increasing width of the ribbon. Armchair-edged ribbons (AGNRs) are
also semiconducting with direct band gaps.\cite{son} Density
functional theory (DFT) calculations predict a high density of
states around the valence and conduction band edges of ZGNRs,
which derive from the states that are localized at the edges of
the ribbon, and lead to non-zero magnetic moments on the carbon
atoms. The existence of such peculiar edge-states in graphitic
materials had been discussed in some earlier
works.\cite{kobayashi,fujita,nakada1,wakabayashi1,nakada2,wakabayashi2}
It has been established that the ground state of a ZGNR is
anti-ferromagnetic with ferromagnetically ordered spin
polarization along the edges.\cite{lee} Due to the fact that the
states near the Fermi level are derived from the edge states and
their linear combinations, external fields have more significant
effects on the edge states. The nature of the interactions between
the magnetized edges of ZGNR was studied, concluding that the main
contribution to the local moments comes from dangling bonds, and
their interaction is determined by tails of the edge-localized
$\pi$ states.\cite{lee} Due to the edge effects, graphene ribbons
show different one dimensional transport properties from those of
carbon nanotubes.

Since GNRs have long spin-correlation lengths and good ballistic
transport characteristics they can be considered as a promising
active material of spintronic
devices.\cite{gambi,dora,vindi,delin} In particular ZGNRs, known
to have large spin polarizations at the opposite edges of the
ribbon, may be utilized to create spin-dependent effects such as
spin polarized currents without the need of ferromagnetic
electrodes or other magnetic entities.

Hydrogen termination of the edge atoms by forming strong $\sigma$
bonds would be important for the structural and electronic
stability of the graphene ribbon. Both first-principles and tight
binding calculations showed that the termination of edges with
hydrogen atoms removes the electronic states related to the
dangling bonds.\cite{kawai} However, there are no qualitative
changes in the electronic structure and the magnetic order of the
ZGNRs with hydrogen atom termination,\cite{kudin} except for a
narrowing of the band gap as we calculated.

Modification of electronic structure by impurities, adatoms and
external fields is another potential of graphene and graphitic
structures for applications in nanoelectronics. Effects on
electronic properties and magnetic behavior of graphene by the
adsorption of foreign atoms has been considered in some previous
works.\cite{lehti, nord, heg, lee2, hod1, gunlycke,kan,sevincli}
Sizeable gap opening by hydrogen adsorption to Stone-Wales defect
sites of graphene has been reported.\cite{dup} In the case of
substitutional boron atom\cite{endo, herna} many of the electronic
properties have been studied and it is suggested that GNRs may be
used as spin filter devices.\cite{martins} Based on the spin
polarized ground state of ZGNRs their possible application as a
spin-valve device was proposed by using tight
binding,\cite{rojas,rycerz} and $\mathbf{k}\cdot\mathbf{p}$
calculations.\cite{brey} It is also predicted by using
first-principles calculations that the electric current flowing on
the ribbon can be made completely spin polarized under in-plane
homogeneous electric fields.\cite{cohen, gunlycke2}

In this study we consider electronic transport properties of short
graphene nanoribbons (graphene flakes), and their
functionalization with vanadium atoms. Electronic and magnetic
properties of various types of nanometer-sized graphene flakes
have been reported
before.\cite{wlwang,shemella,ezawa2,fernandez,jiang,hod2,hod3,rudberg,booth}
We restrict our considerations, however, to rectangular flakes
only, which are finite segments of perfect GNRs. In particular we
calculate spin-dependent transmission spectra of the flakes when
the electrodes make partial contacts along the zigzag rows of the
flakes. The geometrical asymmetry of the contacts lead to a
polarization in the spin states of transmitted electrons.
Depending on the contact geometry and the electrode thickness the
spin polarization of the transmission shows various forms around
the Fermi level, including perfect (100\%) polarization at certain
energy ranges. Then we perform first-principles calculations to
investigate the optimized geometry and electronic properties of
the flakes with adsorbate vanadium atoms. We discuss the binding
sites for single and double adsorbate atoms on the graphene flake.
By using the relaxed geometry of adsorbate-flake system, we
calculate their conductance spectrum by using model metallic
electrodes. The adsorbate vanadium atoms introduce additional
states around the Fermi level modifying the conductance spectrum
of the flake by breaking the spin symmetry of the conduction
electrodes even when the electrodes make uniform contacts with the
flake. We also find traces of Fano resonances in the conductance
spectrum of the graphene flakes arising from the coupling of the
extended states of the quantum channel with the localized states
of the flake, which acts like a scattering quantum dot. A local
density of states (LDOS) analysis show that the discontinuity
features with sharp peaks and dips in the conductance spectra are
associated with perfectly localized states of the structure in the
electrode-flake-electrode geometry.

\section{Calculation Methods}

We performed first-principles total energy calculations to obtain
electronic structure and equilibrium geometries of rectangular
graphene flakes, with and without adatoms, based on the
pseudopotential density-functional theory.\cite{kohn} The
spin-dependent exchange-correlation potential is approximated
within the generalized gradient approximation\cite{perdev} (GGA).
The software package Atomistix ToolKit\cite{atk} (ATK), which
employs local numerical basis orbitals and nonequilibrium Green's
function formalism to calculate quantum conductance in
electrode-device-electrode geometry, has been used in all total
energy and transport calculations.

Geometry optimizations of graphene flakes have been done by
relaxing all atomic positions in supercell geometries with
sufficient vacuum spaces (10 \AA~ minimum) to prevent the
interactions with periodic images of the structure. The criteria
of convergence for total energy and Hellman-Feynman forces were
$10^{-4}$ eV and 0.005 eV/\AA, respectively. The electrostatic
potentials were determined on a real-space grid with a mesh cutoff
energy of 150 Ry. Double-zeta double-polarized basis sets of local
numerical orbitals were employed which gives consistent results
with previous calculations\cite{durgun} that used plane-wave basis
sets. In cases of extended nanoribbon calculations the Brillouin
zone has been sampled with (51,1,1) points within the
Monkhorst-Pack k-point sampling scheme.\cite{monk}

\begin{figure}
\includegraphics[scale=0.17]{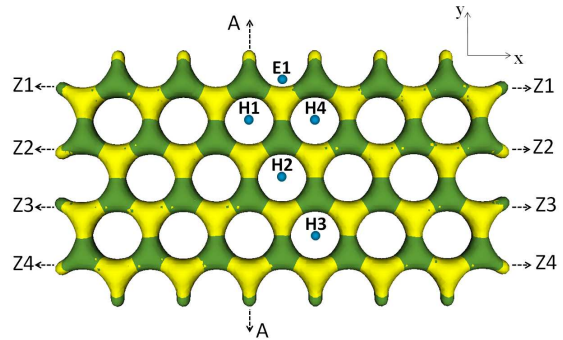}
\caption{(Color online) Geometry and spin-dependent charge density
of the graphene flake cut from 4-ZGNR. The edges are saturated
with hydrogen atoms. Green (dark) and yellow (light) regions
denote the local majority spin-type of the charge density.
Possible adsorption sites of adatoms are labelled as E1, H1--H4.
Electrodes (carbon chain or gold bar) are attached to the flake
along the directions A, Z1--Z4.}\label{zgnr4-vanadium}
\end{figure}

The geometric structure of the graphene flake considered in this
study is shown in Fig.~\ref{zgnr4-vanadium}. It is a finite
segment of 4-ZGNR structure having 4 zigzag rows of carbon atoms
along its axis. The sides along the cut direction have armchair
shape. All edge carbon atoms have been saturated with hydrogen
atoms for a better structural and electronic stability of the
flake. Possible adsorption sites of adatoms have been denoted in
the figure. The magnetic moments of the atoms and the structure as
a whole have been calculated by using Mulliken analysis.

The spin-dependent transport properties of the graphene flake were
calculated based on non-equilibrium Green's function formalism as
implemented in ATK. Ballistic conductance spectrum of the flake is
given within Landauer-B\"{u}ttiker formalism as
\begin{equation}
 G(E)=G_0 \sum_\sigma T_\sigma (E)
\end{equation}
where $G_0=e^2/h$ is the conductance quantum and $T_\sigma$
($\sigma=\textrm{up, down}$) are the spin-dependent transmission
spectra. For each spin state $\sigma$ the transmission
probabilities are calculated as
\begin{equation}
T_\sigma(E)=\textrm{Tr}[\Gamma_L G^r \Gamma_R G^a]_\sigma,
\end{equation}
where $G^r$ and $G^a$ are the retarded and advanced Green's
functions, $\Gamma_L$ and $\Gamma_R$ are the contact broadening
functions associated with the left and right electrodes,
respectively. The broadening functions are anti-Hermitian
components of the self-energy terms of the contacts, $\Sigma_L$
and $\Sigma_R$, which depend on the electrode's surface Green's
function and the contact-molecule (graphene flake) bonding:
$\Gamma_{L,R}=i(\Sigma_{L,R}-\Sigma^\dagger_{L,R})/2$. The
spatially-separated spin-states of the graphene flake are
anticipated to break the spin symmetry in these conductance
calculations when contacts to the electrodes are not symmetric at
the armchair edges.

In order to exploit spin-dependent characteristics of the graphene
flake the structure needs to be partially contacted with thin
metallic electrodes. We used linear carbon atomic chains as model
electrodes. Carbon atomic chains which are known to be
metallic\cite{tongay} are expected to make reasonably good contact
with the flake. The possible contact sites of the electrodes are
denoted in Fig.~\ref{zgnr4-vanadium}. Semi-infinite carbon atomic
chains have been rigidly attached to the flake at one or more
sites along the A, Z1--Z4 directions after removing the hydrogen
atoms at the contact sites. Optimum electrode-flake distances were
obtained by relaxation of the supercell consisting of the flake
and 4-buffer carbon atoms per each carbon chain attached on both
sides of the flakes.

As an alternative to carbon-chain electrodes we also considered
gold electrodes to test the robustness of our conductance
calculations. The gold electrodes were devised by taking a finite
cross-section bar of atoms oriented in the [001] direction of fcc
crystal. The semi-infinite gold bars have a unit cell of nine
atoms in two layers of four and five atoms. We have added single
gold atoms to the ends of the electrodes to get atomically sharp
tips. Contact regions of the electrodes to the flake were
geometrically optimized.

\section{Results and Conclusions}

For the generation of spin polarized currents using a graphene
flake, formation of spin-ordered edge-localized  states found
along the zigzag edges is the key mechanism. Therefore, we first
calculate the ground state magnetic ordering in ZGNR structures
and compare it to that of the flake. In their ground state, all
the carbon atoms of ZGNRs have antiparallel magnetic moments with
their nearest neighbors, and their magnitudes rapidly decrease as
we go away from the edges. For instance, according to the Mulliken
population analysis moments of the atoms in the edge zigzag row of
a 4-ZGNR alternate between 1.21$\mu_{\textrm{\tiny{B}}}$ and
-0.10$\mu_{\textrm{\tiny{B}}}$ per atom. In the second zigzag row
the corresponding values are 0.19$\mu_{\textrm{\tiny{B}}}$ and
-0.07$\mu_{\textrm{\tiny{B}}}$. Relatively high magnetic moments
are found on the atoms that are under-coordinated. Upon saturation
of these carbons with hydrogen atoms the moments on the edge
carbon atoms are reduced but the qualitative properties of the
magnetic ordering of the ZGNRs persisted. For comparison, in the
hydrogen-saturated 4-ZGNR the edge-row carbon atoms have
alternating moments of 0.26$\mu_{\textrm{\tiny{B}}}$ and
-0.06$\mu_{\textrm{\tiny{B}}}$. In the second row the sequence of
moments reduce to 0.07$\mu_{\textrm{\tiny{B}}}$ and
-0.04$\mu_{\textrm{\tiny{B}}}$. The hydrogen atoms too attain
minute moments of about 0.01$\mu_{\textrm{\tiny{B}}}$, aligned
antiparallel to those of the carbon atoms they are attached to.
The spin-polarized edge structure of hydrogenated ZGNRs persists
for wider ribbons, even with a slight increase in the moments of
edge atoms. Hydrogenization also decreases the energy band gap of
the $r$-ZGNRs. For instance, we calculated that the band gap
values for $r=$3, 4, 5 reduce from 1.25, 1.15, 1.00, to 0.70,
0.65, 0.60 in units of eV, respectively.

Geometric and electronic structures of finite segments of ZGNRs,
namely rectangular graphene flakes, are quite similar to those of
extended ribbons. Fig.~\ref{zgnr4-vanadium} shows the particular
graphene flake (C$_{52}$H$_{20}$) considered for the calculations
in this study. Additional armchair edges of the flakes are also
saturated with hydrogen atoms. Spin-unrestricted ground state of
the flake corresponds to the antiferromagnetic spin coupling
between the zigzag edges of the flake. The total energy of this
state is 0.31 eV lower than that of the ground state of a
spin-restricted calculation, which is consistent with the
calculations on extended ZGNRs.\cite{cohen} We find that away from
the armchair edges of the flake the magnetic moments of the
zigzag-edge carbon atoms are close to that of infinite ZGNR
structure, but tend to decrease toward the armchair edges. For
instance, the moments of the six carbon atoms located at the
upmost zigzag edge of the flake shown in Fig.~\ref{zgnr4-vanadium}
are 0.09, 0.21, 0.29, 0.29, 0.21, 0.09 in units of
$\mu_{\textrm{\tiny{B}}}$. The magnetic moment profile of the
carbon atoms rapidly converge to the ribbon values for longer
flakes.


We calculate the spin-dependent ballistic conductance of the
graphene flake when contacted with metallic electrodes. Since zero
bias conductance and transmission probability spectra are related
by a simple multiplicative factor in Eq. 1, we frequently use
these terms as equivalent quantities in the following discussions.
By exploiting the spatial separation of spin states at the
opposite zigzag edges of the flake we show that spin-polarized
currents can be achieved without using ferromagnetic electrodes
and magnetic entities. By making partial contacts to the armchair
sides of the flake the spin symmetry in the coupling of the
flake's electronic states to the electrodes is broken.
Consequently the transmission spectrum of the electrons through
the flake becomes dependent on their spin state. For a clear
demonstration of this effect stripped off complications that may
arise from electrode-graphene interactions we chose to use simple,
model electrodes. Carbon linear atomic chains are appealing in
that respect, being metallic, making naturally good contact with
graphene, and having atomically sharp tips.

\begin{figure}
\includegraphics[scale=0.23]{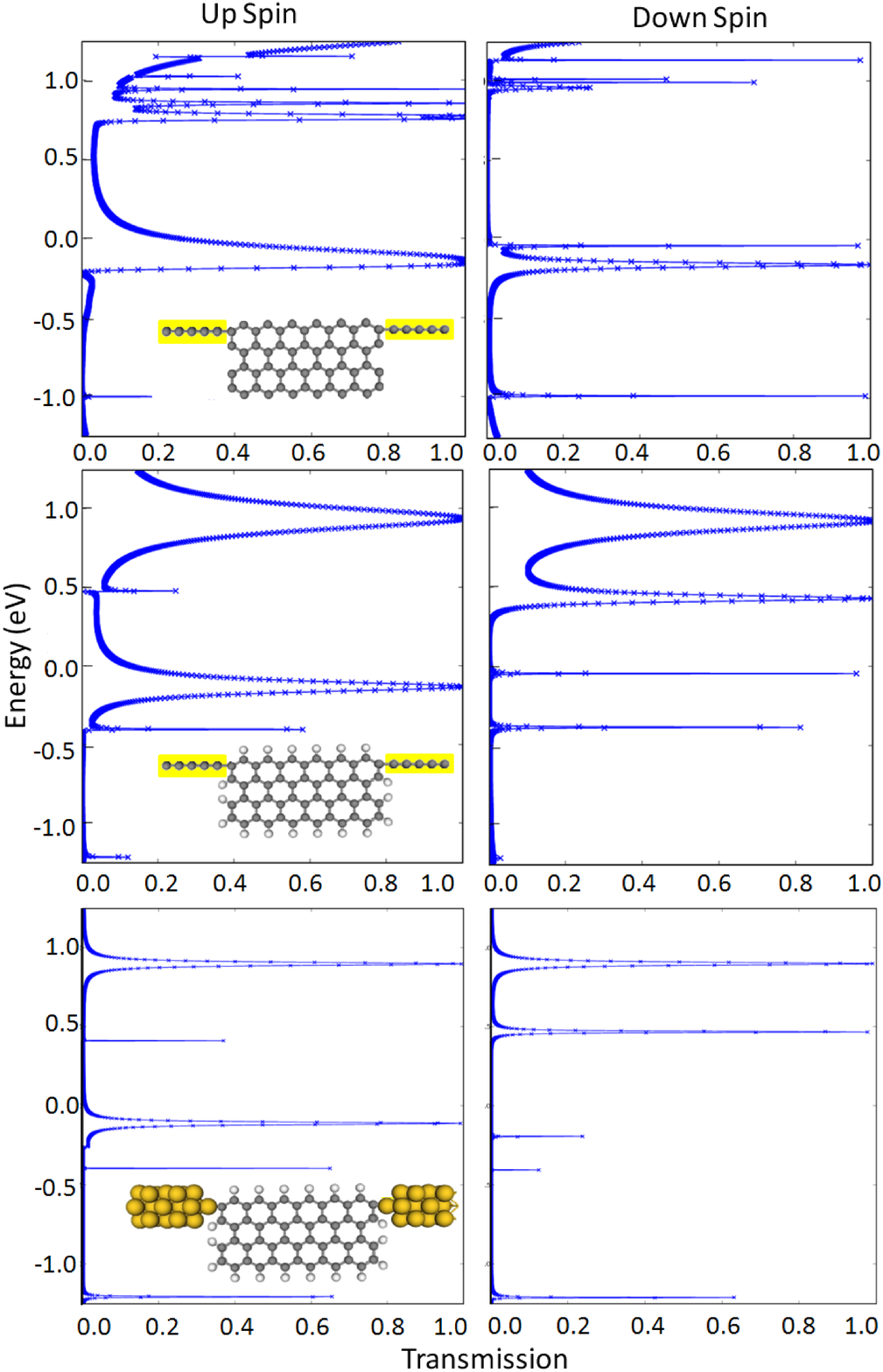}
\caption{(Color online) Spin-dependent transmission spectrum of
the graphene flake without (top panels) and with (middle panels)
hydrogen saturation of the edge atoms. The carbon linear chain
electrodes are connected to the Z1 sites (see
Fig.~\ref{zgnr4-vanadium}). The spectrum when the carbon chain
electrodes are replaced by gold leads are shown in the bottom
panels. Fermi level is set to zero.}\label{passivation}
\end{figure}

In Fig.~\ref{passivation} we plot equilibrium transmission spectra
of the bare and hydrogenated graphene flakes with single carbon
atomic chain electrodes asymmetrically contacted to one of the
edge zigzag rows of the flakes. For comparison, the spectrum of
hydrogenated flake with more realistic gold electrodes is also
presented. Presence of dangling bonds in the bare flake is
reflected in the transmission spectrum as several sharp peaks
about 0.7-1.0 eV above the Fermi level. These peaks are removed
after hydrogenization of the flake, and are replaced by a single
broad peak degenerate for both spin channels. We should note that
within the chosen convention of labelling spin states the
electrodes are contacted at the flake edge where up-spin states
were localized. Therefore, the up-spin states around the Fermi
level couple more effectively with the electrodes leading to
formation of a broader peak for up-spin channel in the
differential conductance spectrum, and consequently the spin
polarization of the current for small bias voltages is expected to
reflect the flake's local majority-spin type of the contact side.
With the gold electrodes, the basic features in the transmission
spectrum remain unchanged except for the transmission peaks being
narrower. The narrower peaks are due to weaker coupling of carbon
atoms to gold electrodes.

\begin{figure}
\includegraphics[scale=0.23]{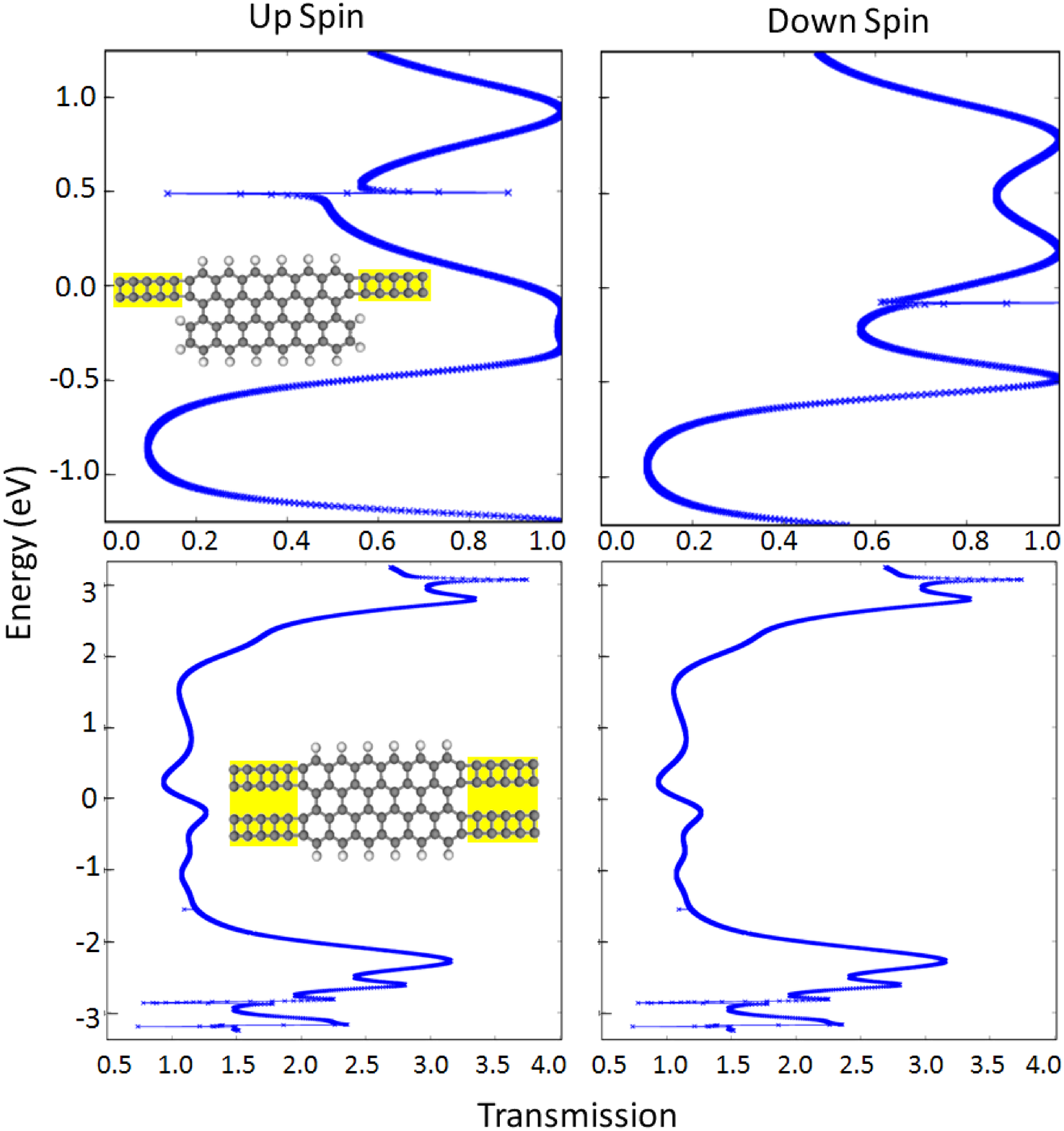}
\caption{(Color online) Spin-dependent transmission spectra of the
hydrogen-saturated graphene flake when the electrodes make partial
contact at the Z1 and Z2 sites (top panels), and when the contacts
are uniform (bottom panels). Fermi level is set to
zero.}\label{z1z2-z1z2z3z4}
\end{figure}

The effect of electrode thickness and its contact location on the
transmission spectrum is further demonstrated in
Fig.~\ref{z1z2-z1z2z3z4}. When the carbon electrodes make a wider
contact with the hydrogenated flake at Z1 and Z2 sites the spin
asymmetry in the conductance spectrum is blurred with broader and
overlapping peaks, but still the up-spin channel has a larger
contribution to the conductance of the flake at energies around
the Fermi level. In case of uniformly contacting electrodes (Z1 to
Z4) the transmission spectrum is identical for both spin channels
as expected. Similarly, conductance of the flake when the
electrodes are contacted along the armchair direction (contact
site A in Fig.~\ref{zgnr4-vanadium}) is spin symmetric (not
shown).


Another mechanism to break spin symmetry in the conductance of
graphene flakes is to introduce magnetic impurities or adatoms.
Here we present a systematic study of adsorption of vanadium atoms
on graphene flake. Single atoms of several elements are adsorbed
on carbon nanotube surfaces with relatively high binding energies
and magnetic moments.\cite{durgun} Similar properties can be
expected for their adsorption on graphene. Vanadium is a suitable
element to consider due to its high binding energy and spin
polarization. The ground state electronic configuration of
vanadium is [Ar]4s$^2$3d$^3$, and it is a chemically active
element with a magnetic moment of 3$\mu_{\textrm{\tiny{B}}}$.


\begin{table}
\caption{Calculated properties of vanadium adatoms on the graphene
flake. See Fig.~\ref{zgnr4-vanadium} for the notation of
adsorption sites. The magnetic order in the flake before the
adsorption is such that the magnetic moments of the carbon atoms
at the top edge are up ($\mu_{\textrm{\tiny{C}}}>0$). In cases of
double adsorbate atoms their locations relative to the flake have
been denoted by top (t) or bottom (b). Magnetic moments of the
vanadium adatoms ($\mu_{\textrm{\tiny{V}}}$) and the total moments
of the structures $\mu_{\textrm{\tiny{tot}}}$ are in units of Bohr
magneton ($\mu_{\textrm{\tiny{B}}}$). $E_\textrm{\tiny{b}}$ is the binding energy per vanadium adatom, and
$d_{\textrm{\tiny{G-V}}}$ is the equilibrium distance of the adatoms to the plane of
graphene flake.} \label{tab:polarize}
\begin{center}
\begin{tabular}{cccccc}
 \hline \hline
 Adsorption site~                 & ~~$\mu_{\textrm{\tiny{V}}}$~~ & ~~$\mu_{\textrm{\tiny{tot}}}$~~& ~~$E_\textrm{\tiny{b}}$ (eV)~~& ~~$d_{\textrm{\tiny{G-V}}}$ (\AA)~~\\ \hline
 \textbf{E1} ($\mu_{\textrm{\tiny{V}}}<0$) &       -3.8    &        -4.9      &    1.59     &   1.94   \\
 \textbf{E1} ($\mu_{\textrm{\tiny{V}}}>0$) &        3.8    &         3.0      &    1.58     &   1.94   \\
 \textbf{H2} ($\mu_{\textrm{\tiny{V}}}<0$) &       -3.9    &        -4.3      &    1.01     &   1.86   \\
 \textbf{H2} ($\mu_{\textrm{\tiny{V}}}>0$) &        3.9    &         4.4      &    1.02     &   1.87   \\
 \textbf{H1-H2} (t-t)                      &     -2.5, 2.0  &        0.0      &   2.09     &1.79, 1.71 \\
 \textbf{H1-H3} (t-t)                      &     3.6, -3.6  &        0.0      &   1.42     &1.77, 1.77 \\
 \textbf{H1-H4} (t-t)                      &     -1.9, 1.8  &        0.0      &   2.41     &1.74, 1.74 \\
 \textbf{H2-H2} (t-b)                      &     3.5, -3.5  &        0.0      &   1.12     &1.75, 1.75 \\
 \textbf{E1-E1} (t-b)                      &     3.7, -3.6  &        0.9      &   1.28     &1.98, 1.96 \\
 \hline \hline
\end{tabular}
\end{center}
\end{table}

We investigated the adsorption of single or double vanadium atoms
on the graphene flake. Possible adsorption sites of vanadium atoms
on the graphene flake are shown in Fig.~\ref{zgnr4-vanadium}. When
there are two vanadium adatoms their location relative to the
flake are denoted by top (t) or bottom (b). Our results are
summarized in Table I. Binding energies per adatom are calculated
by using the expression
$E_\textrm{\tiny{b}}=(E$[G]$+nE$[V]$-E$[G+$n$V])/$n$. Here, $n$ is
the number of adsorbed V atoms, $E$[G] and $E$[V] are the total
energies of the graphene flake (C$_{52}$H$_{20}$) and a free
vanadium atom, respectively. Total energy of the relaxed structure
when foreign atoms are adsorbed on graphene is denoted by
$E$[G+$n$V]. Like most of the 3$d$ transition metal elements (Sc,
Ti, Cr, Mn, Fe, Co) vanadium atoms are adsorbed on hollow sites of
two dimensional graphene sheet. Due to the edge effects, however,
on the graphene flake vanadium atoms are not adsorbed on the H1,
H3, or H4 sites but rather they migrate to the E1 site after
relaxation. We found that vanadium atom can be adsorbed also on
the H2 site, but the binding energy is less than the one at the E1
site. H2-site adsorption is also off-center; equilibrium position
of the vanadium atom tends toward one of the hexagon sides.

\begin{figure}
\includegraphics[scale=0.17]{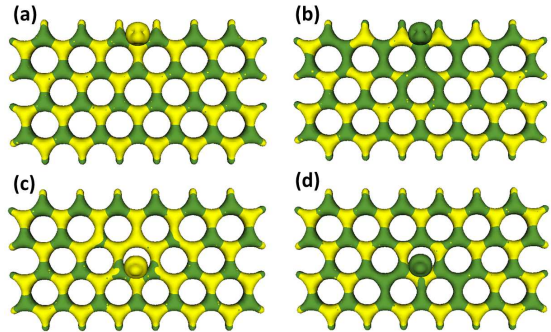}
\caption{(Color online) Majority spin-type of local charge density
of the flake with single vanadium adatoms. The upper zigzag  edge
of the flake was spin-up polarized  before the adsorption. The
vanadium atom has a down magnetic moment in (a) and (c) and an up
moment in (b) and (d). The lowest energy configuration for a
single vanadium atom is shown in (a).}\label{single-van}
\end{figure}

Adsorbed vanadium atoms modify the spin-relaxed charge density of
the flake, however the planer geometric structure of the graphene
flake is quite robust to vanadium adsorption, the displacements of
even the nearest neighbor carbon atoms are less than 0.05 Å after
geometry relaxation. As shown in Table I depending on the initial
spin configurations of the flake and vanadium we identify two
distinct magnetic states with close energies for each adsorption
site of E1 and H2. In its ground state a vanadium atom adsorbed at
the E1 site has a magnetic moment antiparallel with the moments of
the edge carbon atoms as depicted in Fig.~\ref{single-van}(a).

Binding of two adsorbate vanadium atoms on the same side of the
flake is always stronger than that of being on different sides.
For both single-sided and double-sided cases, two vanadium adatom
configurations with antiparallel magnetic moment have larger
binding energies than the cases with parallel magnetic moments.


\begin{figure}
\includegraphics[scale=0.23]{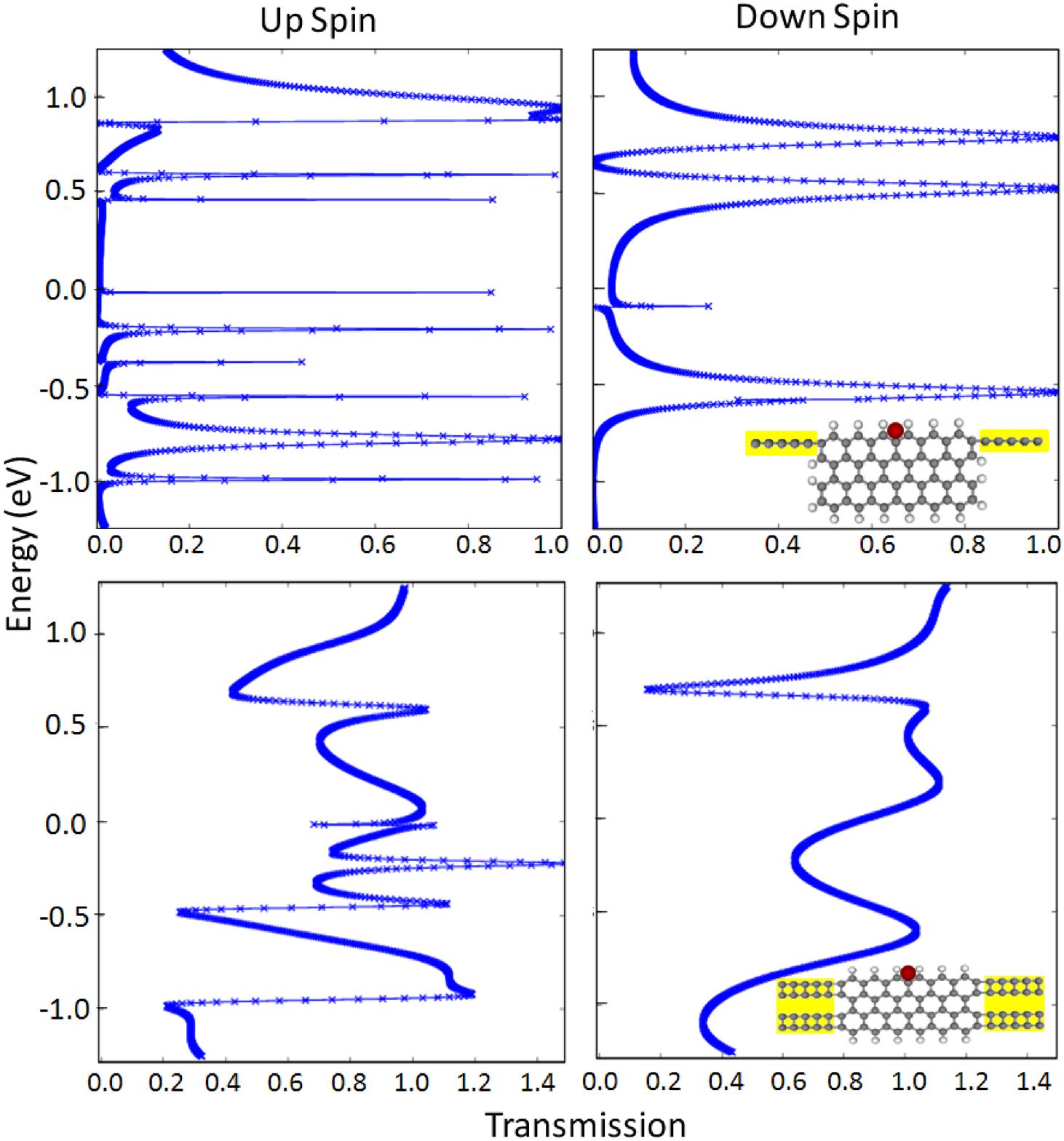}
\caption{(Color online) Spin-dependent transmission spectra of the
hydrogen-saturated graphene flake with an adsorbed vanadium atom
at the E1 site in its ground state. Top panels: The electrodes
make partial contacts at the Z1 sites. Bottom panels: When the
contacts are uniform.}\label{up-vanadium}
\end{figure}

By means of  magnetic dipol-dipol interactions stronger bindings
occur between the edge carbon atoms and vanadium adatom.
Considerable change in the spin charge density of flake with
vanadium adsorbate atoms [see Fig.~\ref{single-van}] also modifies
the conductance spectrum of the structure. In
Fig.~\ref{up-vanadium} it is seen that the adsorption of a
vanadium atom at the E1 site with spin-down magnetic moment
enhances the down-spin transmission channel around the Fermi
level. Spin polarization of the conductance spectrum is present
even when the electrodes make uniform contacts with the flake.


\begin{figure}
\includegraphics[scale=0.25]{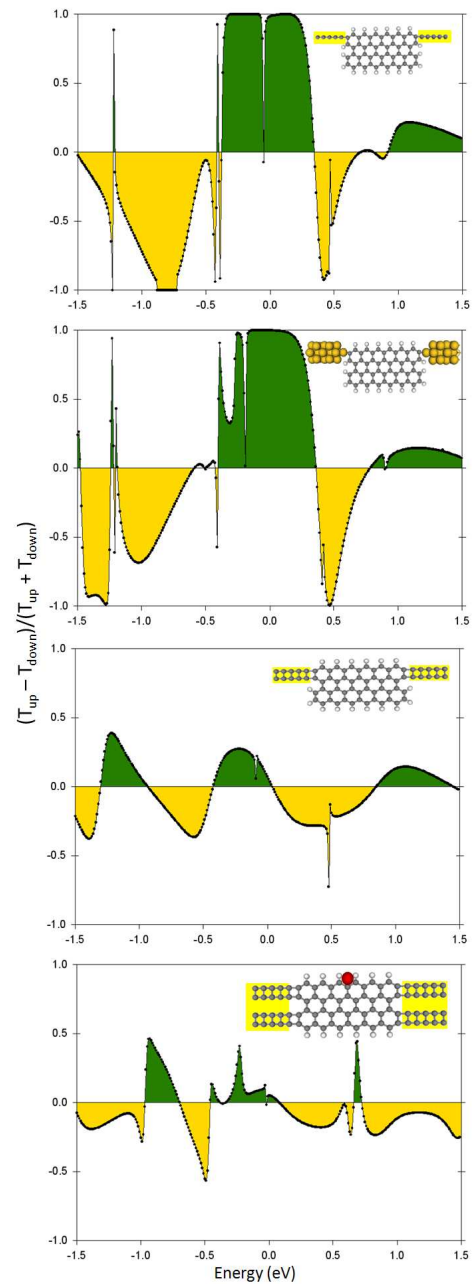}
\caption{(Color online) Spin polarization in the transmission
spectra of the graphene flake. The type and contact geometries of
the electrodes are given as insets. In the lowest panel there is a
single vanadium adatom in its ground state.}\label{polarization}
\end{figure}

In Fig.~\ref{polarization} we present an alternative view of the
spin asymmetry in the conductance spectra by the plots of
energy-dependent relative spin-polarization of the transmission
probabilities,
$(T_{\textrm{\small{up}}}-T_{\textrm{\small{down}}})/(T_{\textrm{\small{up}}}+T_{\textrm{\small{down}}})$,
calculated for four different cases. Partial contacts of the
electrodes and presence of magnetic adatoms both produce spin
polarizations in the spectra. When the flake is contacted from the
Z1 sites either with carbon chain or gold electrodes the
polarization is almost 100\% for an energy window of $\sim$0.8 eV
around the Fermi level. We also see that while uniformly contacted
flake does not exhibit spin polarization, it can be generated by a
single vanadium adatom.

\begin{figure}
\includegraphics[scale=0.2]{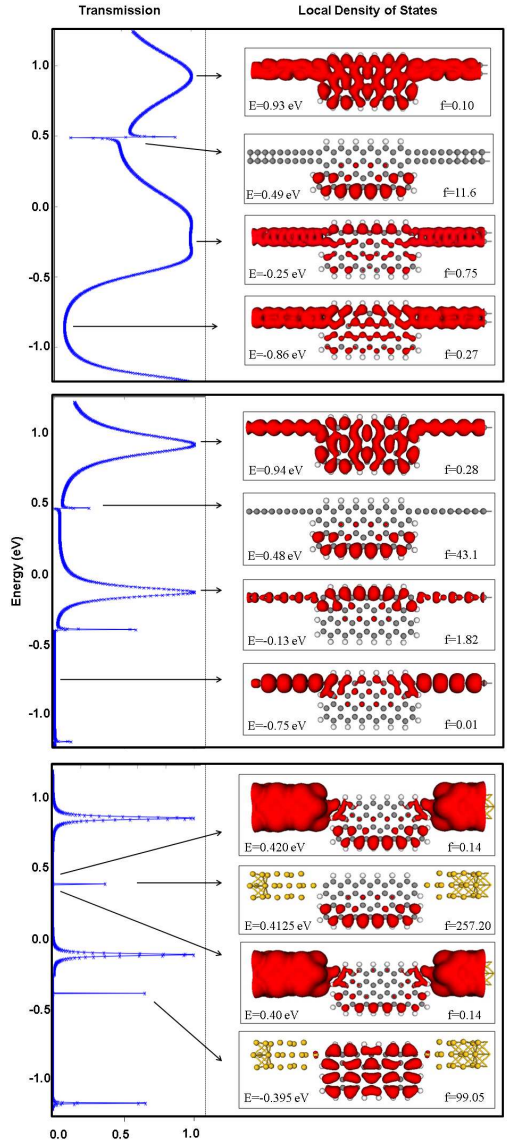}
\caption{(Color online) Local density of states (LDOS) isosurfaces
calculated for particular energy values of the up-spin
transmission spectra of partially contacted graphene flake. The
$f$ value next to each plot is the maximum value of LDOS obtained
in units of states$/($\AA$^3$.eV), and can be taken as a measure
of relative magnitudes among the plots. The plotted isosurfaces
correspond to LDOS values of $f\times 10^{-2}$. }\label{ldos}
\end{figure}

Before we conclude, we would like to discuss several discontinuity
features found in the transmission spectra, a clear example of
which can be seen in top panels of Fig.~\ref{z1z2-z1z2z3z4}. These
are in the form of a pulsative variation with very narrow dips and
adjacent peaks in the transmission function, reminiscent of a
resonant effect. Similar resonances have been recently reported in
the transmission spectra of carbon nanotubes with a single iron or
vanadium adsorbed atom on them.\cite{furst} The underlying
mechanism of these variations in the transmission function is the
Fano resonances of localized states of the structure with extended
band states of the electrodes with matching symmetry. The origin
of the localized states contributing to the resonances can be
either the localized $d$ states of the adatoms as in
Ref.~\onlinecite{furst} and in the vanadium adatom parts of the
present study, or more generally they can be the localized states
of a quantum dot coupled to the continuous states of a quantum
channel.\cite{xu} For instance, in formation of the resonances
found in Fig.~\ref{z1z2-z1z2z3z4} the graphene flake acts like a
scattering quantum dot coupled to the quantum channel of
carbon-chain electrodes.

We infer about the character of these resonances by calculating
local density of states (LDOS) of the structure at resonance and
off-resonance energies of the transmission spectra. In
Fig.~\ref{ldos} we present LDOS isosurface plots for up-spin
transmission spectra of three particular contact geometries. LDOS
are perfectly confined at the flake region only at resonance
energies. At off-resonance energies we always find contribution of
extended states of the metallic electrodes.

In summary, we have performed an \textit{ab initio} study of
spin-resolved conductance spectra of rectangular graphene flakes.
We have shown that spin-polarized currents can be obtained by
making the electrodes contact the flake partially along the
zigzag-edge directions of the flake, or by introducing
transition-metal adatoms such as vanadium to alter the symmetry in
spin-dependent scattering rates of the transmitted electrons. In
the particular model system considered in this study (a
C$_{52}$H$_{20}$ flake cut from 4-ZGNR) a 100\% spin-polarization
in the conductance spectrum was achieved for both model carbon
atomic chain and more realistic gold bar electrodes. Traces of
Fano resonances have been found in the transmission functions. We
believe that the qualitative features of producing spin-polarized
currents in graphene flakes can be verified experimentally with
flakes of larger sizes and with realistic electrodes. An
understanding and realization of spin-dependent effects in
graphitic structures are essential toward the development of
graphene-based spintronic devices.

\begin{acknowledgements}
This work was supported by T\"{U}B\.{I}TAK under Grant No.
106T597, and through TR-Grid e-Infrastructure Project. RTS
acknowledges financial support from Turkish Academy of Sciences
(T\"{U}BA) GEB\.{I}P program.
\end{acknowledgements}

\end{document}